\PassOptionsToPackage{caption=false,font=footnotesize}{subfig}
\documentclass[12,journal]{IEEEtran}
\usepackage{amsmath,amsfonts}
\usepackage{algorithmic}
\usepackage{algorithm}
\usepackage{array}
\usepackage[caption=false,font=normalsize,labelfont=sf,textfont=sf]{subfig}
\usepackage{textcomp}
\usepackage{stfloats}
\usepackage{url}
\usepackage{verbatim}
\usepackage{graphicx}
\usepackage{cite}

\usepackage[caption=false,font=footnotesize]{subfig}
\begin{document}
	

\title{Spectral-Domain Spreading via Hadamard Transform for Robust Downlink Non-Orthogonal Multiple Access}
\author{    Yaakoub Berrouche and Theodoros A. Tsiftsis

    \thanks{Yaakoub Berrouche is with the Department of Electronics, Ferhat Abbas University of Setif 1, Setif, Algeria. Email: berrouchey@univ-setif.com, berrouchey@gmail.com}

}

\maketitle

\begin{abstract}
Non-orthogonal multiple access (NOMA) systems allowing multiple users sharing the same resource block offer significant gains in spectral efficiency which can enable the required massive access in future wireless systems. However, they face several challenges due to their sensitivity to power allocation coefficients, fading effects, and imperfect channel state information (CSI). To address these limitations, this paper proposes Hadamard-NOMA, an approach leveraging the Hadamard Transform (HT) at the source level prior to modulation. By introducing HT, the system mitigates the adverse impact of fading and CSI imperfections, reducing bit error rates (BER) and enhancing overall system reliability. Theoretical analysis and Monte Carlo simulations validate the effectiveness of this technique, demonstrating robust NOMA transmission in dynamic wireless environments. The proposed method offers a promising solution for next-generation wireless networks, ensuring more reliable performance under diverse transmission conditions. Simulation results confirm analytical predictions, demonstrating significant performance improvements over state-of-the-art T-NOMA and Usman-NOMA schemes. Specifically, for the Near user, a gain of 15 dB is achieved at a Bit Error Rate (BER) of $10^{-2}$, while the Far user benefits from a 10 dB gain at a BER of $10^{-1}$. Compared to Usman-NOMA, the proposed method provides an improvement of 15 dB for the Far user at BER $10^{-1}$. Additionally, in a two-user scenario with imperfect Successive Interference Cancelation (SIC), user 1 requires an SNR at least 14 dB lower than user 2 to achieve a BER of $10^{-3}$. These findings highlight the effectiveness of applying HT at the source stage, significantly mitigating CSI errors and making NOMA more resilient for next-generation wireless networks.

\end{abstract}




\IEEEpeerreviewmaketitle

\section{Introduction}

Future wireless networks must be able to support massive access to meet ever-growing demand due to the exponentially increasing number of devices. However, this is not an easy task since the wireless resources (e.g., bandwidth) are limited. In this context, non-orthogonal multiple access (NOMA) is arising as a promising solution.  Unlike conventional techniques such as time division multiple access (TDMA) and frequency division multiple access (FDMA), NOMA allows multiple users to share the same frequency channel simultaneously, significantly enhancing spectrum efficiency while meeting quality of service (QoS) requirements \cite{dai2015non,liu2017nonorthogonal,10615971}. To be able to operate effectively and to ensure reliable communication NOMA needs three key players as follows. \textit{i)} Successive interference cancellation (SIC), where users decode their own data while subtracting interference from others. \textit{ii)} Efficient power allocation (PA), where power levels are assigned based on channel conditions. \textit{iii)} Superposition coding, which combines multiple users' data into a single transmitted signal. By optimizing resource utilization and ensuring fair service distribution, NOMA enhances connectivity and reliability in next-generation wireless networks.

Traditional NOMA (T-NOMA)  achieves high spectral efficiency, but it may require perfect channel state information (CSI) and an impairment-free system. The performance of NOMA is highly affected by CSI errors or system malfunctions, particularly for users who are far from the base station. Moreover, this performance loss gets worse by poor channel conditions and the number of users per cluster which is limited by the cascading error impact caused by SIC. One of the primary challenges for NOMA is ensuring robustness, necessitating the development of sturdier strategies. Nevertheless, due to non-convex optimization obstacles, existing approaches often involve significant computational complexity and offer only limited enhancements in performance. 
In \cite{beddiaf2022unified}, cooperative NOMA (CNOMA), a variant of NOMA, is investigated under hardware impairments at the transceivers and CSI errors. The authors derive expressions for the bit error rate (BER) and reveal that, under high levels of hardware impairments, NOMA can outperform CNOMA. They also highlight the critical role of PA in preventing error propagation, emphasizing that carefully selected PA can substantially enhance performance. In \cite{xie2022robust}, the focus shifts to robust resource allocation (RA) in NOMA-based heterogeneous networks (HetNets) under imperfect CSI. The study addresses a resource allocation problem aiming to maximize energy efficiency, presenting valuable insights into the interplay between resource management and system robustness. These works collectively underscore the importance of addressing CSI errors, hardware impairments, and power allocation to enhance the reliability and robustness of NOMA systems in practical scenarios.Although recent studies have started tackling issues such as CSI errors, improper SIC, and hardware imperfections, research into advancing NOMA performance still remains as an open problem.
One way of improving the resilience of NOMA against SIC errors is the integration of multiple description coding (MDC). Several studies have explored MDC’s potential to enhance reliability and performance \cite{wu2019video, dani2022resource, tang2020mdc}. In \cite{tang2020mdc}, the authors developed a hybrid MDC-NOMA scheme for image transmission, combining MDC principles with NOMA to improve robustness. 
The simulations demonstrated that even if a user decodes only one of the multiple descriptions, the transmitted image can still be successfully reconstructed, highlighting the scheme’s efficiency and resilience. While MDC-based NOMA systems \cite{wu2019video, dani2022resource, tang2020mdc} exhibit improved robustness, they do not fully exploit the correlation properties of MDC due to their application at the symbol level.

As a subset of MDC, the Hadamard transform (HT) has been also explored in NOMA systems in \cite{usman2018joint,8441999,bouslam2024noma}. In \cite{usman2018joint}, the authors applied HT at the transmitter after modulation. However, this approach introduced unnecessary complexity by distributing symbols with an unclear physical meaning and employing complex numbers. 
In \cite{8441999}, the authors employed HT post-modulation to address the peak-to-average power ratio (PAPR) in multi-carrier systems by focusing power optimization. Similarly, in \cite{bouslam2024noma}, the authors applied HT post-modulation in a four-user NOMA system and analyzed the sum-rate of NOMA comparing its OMA counterpart. However, both studies failed to explore the effect of HT on the BER performance of NOMA systems and increase the computational complexity due to post applications. On the other hand, applying HT before modulation at the user data level streamlines the process, preserves the physical meaning of the data, and aligns with MDC theory, as supported by earlier research (e.g., \cite{berrouche2014improved}). This approach ensures theoretical soundness and enables a reliable application of HT to enhance NOMA systems.



To address the aforementioned limitations in the existing literature, this study proposes integrating HT into the NOMA framework, creating the H-NOMA system. HT enhances correlation among multiple descriptions and spreads user data, improving resilience against CSI errors, SIC errors, and hardware impairments. Main contributions of this work are summarized as follows.

\begin{itemize}
\item We propose a novel HT-based NOMA framework grounded in theoretical principles. We present and rigorous mathematical analysis to optimize H-NOMA systems under both perfect and imperfect successive interference cancellation (SIC) conditions.
\item We present extensive simulation results for the H-NOMA system under different fading conditions (e.g., Rayleigh and Nakagami-m) to demonstrate improved key performance metrics, including BER reduction and SNR improvements while strengthening NOMA's robustness against channel impairments such as multi-path fading and poor CSI
\item We provide practical guidelines for integrating HT into next-generation 5G and 6G wireless networks to enhance throughput, reliability, and overall performance in real-world scenarios such as image processing, broadening its impact beyond conventional telecommunications.
\end{itemize}
The remainder of the paper is structured as follows. Section \ref{systemModel} presents the system model, describing the communication framework, user grouping, and power allocation. We review the principles of traditional NOMA, focusing on power domain multiplexing and SIC limitations. Likewise, we introduce the Hadamard transform-based NOMA (H-NOMA) scheme and discuss its advantages in mitigating interference. Moreover, we analyze SIC with imperfect channel conditions in H-NOMA and its impact on performance. 

Section \ref{analyticalAnalysis} provides a theoretical framework for evaluating system performance, including error probability analysis. We examines the effects of SIC errors, considering different interference levels and imperfect cancellation. Likewise, we presents a mathematical formulation of the BER, incorporating the impact of fading and SIC errors. 

Section \ref{simulationResults} discusses numerical simulations validating the proposed model through BER performance comparisons with traditional NOMA. We presents BER results under different system conditions in order to demonstrate scalability. likewise, we investigates BER performance under imperfect SIC conditions for two users, highlighting the impact of residual interference. Section \ref{realTimeApplication} explores the feasibility of deploying H-NOMA in practical applications, such as image and data transmission. Finally, Section \ref{conclusion} summarizes key findings and suggests potential directions for future research.

\textit{Notation:} Lowercase bold symbols denote vectors, and uppercase bold symbols denote matrices. We use $\mathbb{C}$ for the set of complex numbers, $\lfloor\cdot\rfloor$ for the floor operator, $\mathbf{I}_N$ for the $N \times N$ identity matrix, and $\mathbf{0}$ for a zero vector of appropriate dimension. The operators $(\cdot)^T$, $(\cdot)^*$, and $(\cdot)^H$ represent transpose, complex conjugate, and conjugate transpose, respectively; $\mathfrak{R}(\cdot)$ denotes the real part, and $\mathbb{E}[\cdot]$ denotes expectation. The Euclidean (or $\ell_2$) norm of $x \in \mathbb{R}^n$ is $\|x\|_2$, and the Frobenius norm of $A \in \mathbb{R}^{m \times n}$ is $\|A\|_F$. The operators $\mathbf{A} \odot \mathbf{B}$ and $\mathbf{A} \oslash \mathbf{B}$ denote the Hadamard (elementwise) product and division, and $\mathbf{A}^{\circ 2}$ is the elementwise square of $\mathbf{A}$. For a vector $\mathbf{v}$, $\operatorname{diag}(\mathbf{v})$ is the diagonal matrix with entries of $\mathbf{v}$ on its main diagonal, while $\operatorname{diag}(\mathbf{M})$ returns the vector containing the diagonal entries of $\mathbf{M}$. Complex Gaussian random variables are written as $\mathcal{CN}(\mu,\sigma^2)$ with mean $\mu$ and variance $\sigma^2$. The $n$‑th diagonal element of a matrix $\mathbf{A}$ is denoted by $(\mathbf{A})_n$. The elementwise absolute value of a vector $\mathbf{a}$ is $|\mathbf{a}|=\left(\left|a_0\right|,\ldots,\left|a_N\right|\right)^T$. The entry $m_{ij}$ refers to the element in the $i$‑th row and $j$‑th column of $\mathbf{M}$, and $\mathbf{e}_i$ denotes the $i$‑th canonical basis vector in $\mathbb{R}^N$. A complete list of notation is summarized in Table I.

\section{System Model}
\label{systemModel}
\begin{figure}
    \centering
    \includegraphics[width=9cm]{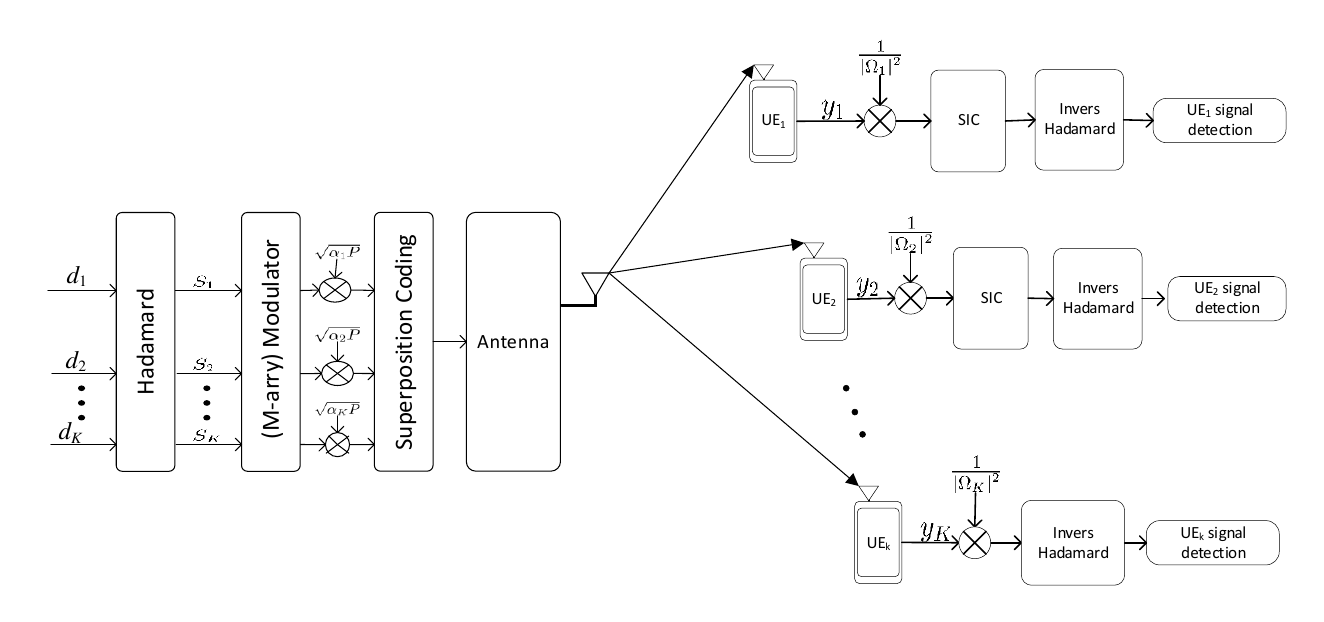}
    \caption{System model of downlink NOMA.}
    \label{fig:system_model}
\end{figure}

This paper considers downlink NOMA transmissions where a base station (BS) serves $K$ single-antenna users (denoted as $\mathcal{K}=\{1,2,...,K\}$) where the channel condition is sorted from the best to the worst as shown in Fig.~\ref{fig:system_model}. In traditional orthogonal schemes, the BS uses TDMA to serve $K$ users within the assigned resources. In conventional NOMA schemes, the BS employs superposition coding to serve $K$ users on the same time-frequency resources, the user with the highest power allocation decodes directly from the received signal (treat the intended signal from other users as noise), and user $i=\{2,3, ..., K\}$ performs $j=\{1,2, ..,i-1\}$ SIC to remove the NOMA interference before decode its own signal. 

\subsection{Traditional NOMA}
\vspace{-0mm}
A downlink NOMA system with a single base station (BS) and $N$ users is considered. The users are arranged according to their channel quality so that $\left|g_{N}\right| >... > \left|g_{2}\right| > \left|g_{1}\right|$, where $g_k \in \mathbb{C}$. Every user adheres to their power limits while sharing the same time-frequency resources, and the BS is fully aware of the CSI. 
The message intended for the $k^{th}$ user ($U_{k}$) is denoted as $d_{k}\in \{0, 1\}$, for $1\leq k \leq N$, where $N$ denotes the total number of users. The input data vector for all the users is represented as a vector,
\begin{equation}
\mathbf{d} = [d_{1}, d_{2}, \dots, d_{N}]^T \in \{0,1\}^{N \times 1}.
\label{dvector}
\end{equation}
For each user $U_{k}$. Quadrature amplitude modulation (QAM) is applied to modulate ${d}_k$ resulting in the modulated signal $x_k$. For better decoding performance, users with weaker channels receive higher power during the power-domain multiplexing operation, while users with stronger channels receive lower power. All users receive a superimposed signal from the base station, which is provided by:
\begin{equation}
x=\sum_{k=1}^{N} \sqrt{P_{k} \alpha_{k}} x_{k}, 
\end{equation}
\noindent where $P_k = \alpha_k P_s $ represents the power allocated to the $k$-th user, while $x_k$ denotes the modulated complex data symbol derived from the transformed data $w_{k}$. Additionally, $\alpha_{k} \in \mathbb{R}$ is the power allocation coefficient assigned to $U_{k}$, which is subject to the constraint
$\sum_{k=1}^{N} \alpha_{k}=1, \:\:\: 0 \leq \alpha_k \leq 1, \:\:\:,
\mathbb{E}\left[\left|x_{k}\right|^{2}\right]=1$.
This ensures that the entire transmission power, $P_{s}$, is distributed among users in a suitable manner, giving users who are facing extreme fading conditions a larger power allocation. 
The users' channel gains are utilized to determine the power allocation coefficients, following the order $\alpha_{1}> \alpha_{2}> \dots >\alpha_{N}$. 

At the receiver front-end, for flat fading channels, the baseband signal received by user $U_{k}$ is given by
\begin{equation}
y_{k}=g_{k}x+n_{k}, \quad 
\end{equation}
\vspace{-3mm}

\noindent where $g_{k}$ represents the complex channel fading coefficient between the BS and the $k^{th}$ user, 
capturing the effects of the channel, and $n_{k}$
 denotes the noise at the receiver, modeled as additive white Gaussian noise (AWGN) with zero mean and variance  $\sigma_{n_k}^{2}$. 

\subsection{Proposed H-NOMA}
\begin{figure}
    \centering
    \includegraphics[width=9cm]{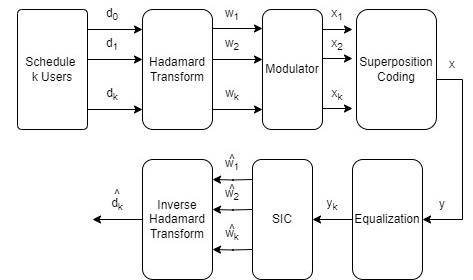}
    \caption{System model of H-NOMA.}
    \label{fig:SystemmodelofH-NOMA}
\end{figure}

In the proposed H-NOMA scheme, we deploy HT before IQ modulation to increase the robustness against channel and hardware impairments. Therefore, the information vector (\ref{dvector}) for each user is first processed through the HT block, as illustrated in Fig. \ref{fig:SystemmodelofH-NOMA}. To generate the H-NOMA symbol, the input data vector (\ref{dvector}) undergoes an HT of order $N$, producing the transformed data vector $\mathbf{w}$ using the Sylvester-Hadamard matrix $\mathbf{H}_N$ as follows, 
\begin{equation}
\mathbf{w}=\mathbf{H}_{N} \mathbf{d} \in \mathcal{M},
\label{vectorw}
\end{equation}
\noindent where $\mathcal{M} \in \{\pm N, \pm (N-1), ...\}^{N \times 1}$. The Sylvester-Hadamard matrix of order $2$ is defined as \cite{1091168},
 \begin{equation}
\mathbf{H}_{2}=\frac{1}{\sqrt{2}} \begin{bmatrix}
1 & 1 \\
1 & -1 \\
\end{bmatrix}.
\end{equation}
For higher-order Hadamard matrices, the recursive construction follows,
\begin{equation}
\mathbf{H}_{N}=\mathbf{H}_{N-1}\otimes \mathbf{H}_{2},
\end{equation}
 
 where $\otimes$ denotes the Kronecker product, and $N$ is a positive integer. These matrices exhibit the unitary property \cite{1091168}. The transformed data vector (\ref{vectorw}) is then affined transformed as $\mathbf{w}' = \mathbf{w}+ m\mathbf{1}$ where $m = N/2$ and $\mathbf{1}$ is all-ones vector of the dimension $N \times 1$. Each $w'_k$ is mapped to a higher-order constellation (e.g., $M$-QAM) to obtain the modulated symbol $x_k$.

\subsection{SIC with Imperfect Channel in H-NOMA}

Given the channel coefficient $g_k$, power allocation coefficient $\alpha_k$, total transmission power $P_s$, and received signal $y_k$ at the $k^{th}$ user, SIC is employed to mitigate interference. Since signals from $(k-1)^{th}$ users interfere with the $k^{th}$ user's received signal, SIC ensures accurate detection by sequentially removing interference. Unlike traditional NOMA, where a user treats lower-power signals as interference, H-NOMA exploits the transformed domain to decode all $N$ users’ signals jointly.  

At the receiver, for $y_{1}$, given its own channel coefficient $g_1$ and power allocation $\alpha_1$, user 1 decodes its own signal using likelihood detection (MLD) according to the following rule,

\begin{equation}
\hat{x}_{1} =arg\displaystyle\max_{\hat{x} \in \mathcal{S}} |y_{1}-\sqrt{P_{s}\alpha_{1}}g_{1} \hat{x} |^{2},
\end{equation}

\noindent where $\mathcal{S}$ denotes the signal constellation set. To obtain user 2's symbol $\hat{x}_2$, user 1’s estimate $\hat{x}_1$ is subtracted from the aggregate received signal $r_{1}$ as follows,

\begin{equation}
\hat{x}_{2} =arg\displaystyle\max_{\hat{x} \in \mathcal{S}} |r_{1}-\sqrt{P_{s}\alpha_{1}}g_{1} \hat{x} |^{2},
\end{equation}

\noindent where, 

\begin{equation}
r_{1}= y_{1}-\sqrt{P_{s}\alpha_{1}}g_{1} \hat{x}_1 .
\end{equation}

The estimated signals are then transformed back:

\begin{equation} \hat{\mathbf{w}} = \hat{\mathbf{w}}' - m\mathbf{1}. \end{equation}

If user 1’s transformed signal $\hat{w}_{1}$ is correctly decoded, user 2’s transformed signal $\hat{w_{2}}$ can be recovered interference-free. Otherwise, errors in user 1’s decoding will propagate, affecting user 2. After recovering the transformed vector $\hat{\mathbf{w}} = [\hat{w}_1, \dots, \hat{w}_N]^T$, the estimated binary data is obtained via inverse HT as such:

\begin{equation}
\hat{\mathbf{d}}=\mathbf{H}_{N}^{-1}\hat{\mathbf{w}}.
\end{equation}

Finally, the estimated user data $\hat{d}_{k}$ is extracted from $\hat{\mathbf{d}}$. Unlike the T-NOMA scheme, where the estimation of $\hat{d}_{k}$ relies solely on $\hat{w}_{k}$, rendering all previously decoded data $\hat{w}_{k-1}$ irrelevant, the proposed H-NOMA scheme utilizes all transformed data elements $\hat{w}_{k}$, $k \in \left\{ 1, 2, ...N\right\}$, in the estimation process, enhancing the overall detection performance.

We consider a wireless communication system where the channel matrix $\mathbf{G}$ remains constant during the transmission of the signal matrix $\mathbf{S}$. The receiver computes an estimate $\hat{\mathbf{G}}$ of the actual channel matrix $\mathbf{G}$. The channel model is given by \cite{borran2009design}:

\begin{equation}
    \mathbf{G} = \hat{\mathbf{G}} + \tilde{\mathbf{G}},
\end{equation}

where $\hat{\mathbf{G}} \sim \mathcal{CN}(0, (1 - \sigma_E^2) \mathbf{I})$ represents the estimated channel known to the receiver, and $\tilde{\mathbf{G}} \sim \mathcal{CN}(0, \sigma_E^2 \mathbf{I})$ represents the unknown channel component, modeled as an additive channel estimation error.

The entries of $\hat{\mathbf{G}}$ and $\tilde{\mathbf{G}}$ are assumed to be independent, which can be achieved using a linear minimum mean square error (LMMSE) estimator. The parameter $\sigma_E^2 \in [0, 1]$ denotes the channel estimation variance per channel coefficient. When $\sigma_E^2 = 0$, the model corresponds to a coherent system, while $\sigma_E^2 = 1$ represents a non-coherent system.

To generate the channel coefficients for far and near users, we use the following expressions:

\begin{align}
    \hat{g}_{\text{1}} &= (1 - \sigma_E^2) \sqrt{d_{\text{Far}}^{-\eta}} \frac{\mathcal{CN}(0,1)}{\sqrt{2}}, \\
    \hat{g}_{\text{2}} &= (1 - \sigma_E^2) \sqrt{d_{\text{Near}}^{-\eta}} \frac{\mathcal{CN}(0,1)}{\sqrt{2}}, \\
    \tilde{g}_{\text{1}} &= (\sigma_E^2 )\sqrt{d_{\text{Far}}^{-\eta}} \frac{\mathcal{CN}(0,1)}{\sqrt{2}}, \\
    \tilde{g}_{\text{2}} &= (\sigma_E^2) \sqrt{d_{\text{Near}}^{-\eta}} \frac{\mathcal{CN}(0,1)}{\sqrt{2}},
\end{align}

where $\mathcal{CN}(0,1)$ represents a standard complex normal variable. The overall channel coefficients for the far and near users are then given by:

\begin{align}
    g_{\text{1}} &= \hat{g}_{\text{1}} + \tilde{g}_{\text{1}}, \\
    g_{\text{2}} &= \hat{g}_{\text{2}} + \tilde{g}_{\text{2}}.
\end{align}

\vspace{-3mm}
\section{Analytical Analysis}
\label{analyticalAnalysis}

\subsection{Analysis of SIC error}
To align with the current understanding of H-NOMA receivers, we simplify the analysis by considering a two-user scenario, i.e., $N = 2$. In a typical two-user H-NOMA system, the base station transmits signals to both users simultaneously, allocating different power levels based on their respective channel conditions.
For two users, where $d_{1}$ represents the far user's data and $ d_{2}$ represents the near user's data, the transformed data symbols are obtained as follows
\begin{equation}
w_{1}=d_{1} + d_{2}, \quad
w_{2}=d_{1} - d_{2}.
\end{equation}
The transformed signals $w_{1}$ and $w_{2}$ are are mapped to BPSK to obtain the transmitted symbols $x_1$ and $x_2$. Consequently, the H-NOMA transmitted signal is formulated as,
\begin{equation}
x=\sqrt{P_1}x_{1}+\sqrt{P_2} x_{2}.
\end{equation}
At the near user, the received H-NOMA signals is given by
\begin{equation}
r=g_{2}\sqrt{P_1}x_{1}+g_2\sqrt{P_2} x_{2}+n.
\end{equation}
After performing SIC, the estimated symbols are obtained as,
\begin{equation} \hat{x}_1 = \frac{r}{g_2\sqrt{P_1}}, \quad
\hat{x}_2 = \frac{r - g_2\sqrt{P_1}x_1}{g_2\sqrt{P_2}}. \end{equation}
Accordingly,
\begin{equation}
\hat{x}_{1}=x_1
;
\hat{x}_{2}=x_2+\frac{n}{g_2\sqrt{P_2}}
\end{equation}
Similarly, we derive
\begin{equation} \hat{x}_1 + \hat{x}_2 = x_1 + x_2, \quad
\hat{x}_1 - \hat{x}_2 = x_1 - x_2 + \frac{n}{g_2\sqrt{P_2}}. \end{equation}
Subtracting these two equations, we obtain the symbol estimation in H-NOMA
\begin{equation}
\hat{x}_1=x_1-\frac{n}{2g_2\sqrt{P_2}.}
\end{equation}
The noise term $\frac{n}{2g_2\sqrt{P_2}}$ introduces deviations in $\hat{x}_1$, which the receiver utilizes to estimate the near-user data
$\hat{d}_1$. In contrast, in NOMA, the estimation follows
\vspace{-2mm}
\begin{equation}
\hat{x}_1= x_1+\frac{n}{g_2\sqrt{P_2}}.
\end{equation}
This result implies that H-NOMA effectively mitigates noise interference by partially canceling it, whereas NOMA does not, potentially leading to reduced accuracy in decoding the near-user signal $d_1$.
Similarly, a higher channel gain $h_2$ (stronger channel) reduces the impact of noise on the decoded signal in H-NOMA by a factor of two compared to NOMA. This enhances the reliability of the decoded signal, ensuring that the estimated near-user data $\hat{d_1}$ closely matches the original data signal $d_1$, leading to more accurate decoding. For the far user, the received H-NOMA signal is given by
\begin{equation}
r=g_{1}\sqrt{P_1} x_{1}+g_1\sqrt{P_2} x_{2}+n
\end{equation} 
After performing SIC, the estimated symbols are
\begin{equation}
\hat{x}_{1}=\frac{r}{g_1\sqrt{P_1}}
\quad 
\hat{x}_{2}=\frac{r-g_1\sqrt{P_1}x_2}{g_1\sqrt{P_2}}
\end{equation}
Therefore,
\begin{equation}
\hat{x}_{1}=x_1+\sqrt\frac{P_2}{P_1}x_2+\frac{n}{g_1\sqrt{P_1}}
\quad 
\hat{x}_{2}=x_2+\frac{n}{g_1\sqrt{P_2}}
\end{equation}
Further, we derive
\begin{equation}
\hat{x}_1+\hat{x}_2=x_1+x_2+\sqrt\frac{P_2}{P_1}(x_1-x_2)+\frac{n}{g_1\sqrt{P_1}}
\end{equation}
\begin{equation}
\hat{x}_1-\hat{x}_2=x_1-x_2+\frac{n}{g_1\sqrt{P_2}}
\end{equation}
Adding these two equations, we obtain the symbol estimation in H-NOMA,
\begin{equation}
\hat{x}_1=\frac{1}{2} (2+\sqrt\frac{P_2}{P_1})x_1-\frac{1}{2}\sqrt\frac{P_2}{P_1}x_2+
\frac{n}{2g_1\sqrt{P_1}}+\frac{n}{2g_1\sqrt{P_2}}
\end{equation}
On the other hand, in NOMA, the estimated symbol is
\begin{equation}
\hat{x}_{1}=x_1+\sqrt{\frac{P_2}{P_1}}x_2+
\frac{n}{g_1\sqrt{P_1}}
\end{equation}
H-NOMA improves the accuracy of the far-user decoded signal $\hat{d_1}$ by effectively canceling near-user interference ($x_2$). In contrast, NOMA does not eliminate near-user interference, which may degrade the accuracy of $\hat{d_1}$. Additionally, H-NOMA enhances robustness by amplifying the far-user symbol ($x_1$) nd considering noise contributions from both power levels ($P_1$ and $P_2$). T-NOMA introduces more interference, leading to higher error rates in demodulation and decision-making. However, H-NOMA performs better at mitigating near-user interference, particularly when $P_1 \approx P_2$.
\vspace{-2mm}
\subsection{Bit Error Rate Analysis}






The H-NOMA signal transmitted by the BS is defined as \cite {assaf2020exact}:
\vspace{-2mm}
\begin{equation}
x=\sum_{k=1}^{N} \sqrt\frac{P_{s} \alpha_{k} }{E_k} x_{k} 
\end{equation}

\begin{table}[]
\caption{Abbreviation}
\resizebox{8cm}{!}{
\begin{tabular}{|l|l|}
\hline$B_{k}=\log _{2} M_{k}$ & $L_{1, k}=\left(1-2^{-i}\right) \sqrt{M_{k}}-1$ \\
\hline$\Lambda_{k}=\sqrt{M_{k}}-1$ & $M=M_{1} \times M_{2}$ \\
\hline$v_{k}=\log _{2} \sqrt{M_{k}}$ & $\varphi_{1}=\frac{1}{\sqrt{2}}\left(\sqrt{\alpha_{1}}+\sqrt{\alpha_{2}}\right)$ \\
\hline$\lambda_{i, n, k}=\left[\frac{n 2^{i-1}}{\sqrt{M_{k}}}\right.$ & $\varphi_{2}=\frac{1}{\sqrt{2}}\left(\sqrt{\alpha_{1}}-\sqrt{\alpha_{2}}\right)$ \\
\hline$\epsilon_{k}=\frac{\beta_{n}}{\sigma_{\mathrm{k}}}$ & $g_{k}^{ \pm}(a, b)=\epsilon_{n} \sqrt{q_{k}^{-\zeta}}\left(a \sqrt{\frac{\alpha_{1}}{E_{1}}} \pm b \sqrt{\frac{\alpha_{2}}{E_{2}}}\right)$  \\ \hline
\end{tabular}}
\end{table}

where, 
\begin{equation}
Q\left(g_{1}(a, b)\right)=Q\left(g_{1}^{+}(a, b)\right)-Q\left(g_{1}^{-}(a, b)\right).
\end{equation}
The analytical BER expression for user 1 is given in \cite{assaf2020exact} as derived from [5, (9), (14)]:  
\begin{equation}
P_{b_{1 k}}\!\!=\!\!\frac{1}{\sqrt{M}}\!\! \sum_{i=0}^{L_{1,1}} \sum_{l=0}^{\Lambda_{2}}\!\! D_{1}(k, i) Q\!\!\left(g_{1}^{+}\!\!\left(2 i+1,2 l\!\!-\!\!\sqrt{M_{2}}\!\!+\!\!1\!\!\right)\!\!\right), 
\end{equation}
\noindent where
\begin{equation}
D_{1}(k, i)=(-1)^{\lambda_{k, i, 1}}\left(2^{k-1}-\left\lfloor\frac{i 2^{k-1}}{\sqrt{M_{1}}}+\frac{1}{2}\right\rfloor\right).
\end{equation}







Therefore, the average unconditional BER of user 1 is computed as 
\begin{equation}
\bar{P}_{U_{1}}=\frac{2}{v_{1}} \sum_{k=1}^{v_{1}} \bar{P}_{b_{1 k}}.
\end{equation}
The analytical BER expression for user 2 is provided in \cite{assaf2020exact}, as derived from [5, (19)(20)]:
\begin{equation}
\begin{aligned}
& \bar{P}_{b_{2 k}}\!\!=\!\!\sum_{i=0}^{L_{1,2}} \sum_{l=0}^{2 \Lambda_{1}} \!\!\frac{P_{C}^{+}}{\sqrt{M}} S D_{2}(k, i) D_{3}(k, l) Q\left(g_{2}^{+}\left(c_{k} l, 2 i+1\right)\!\right) \\
& \left.\!\!-\!\!\sum_{i=0}^{L_{1,2}} \sum_{l=1}^{2 \Lambda_{1}} \!\!\frac{P_{C}^{-}}{\sqrt{M}} S D_{2}(k, i) D_{3}(k, l) \times Q\left(g_{2}^{-}\left(c_{k} l, 2 i+1\right)\right)\!\!\right) 
\end{aligned}
\end{equation}
\noindent where $P_{C}^{ \pm}=\left[1-\sqrt{\frac{\bar{\gamma}_{2}^{ \pm}}{\bar{\gamma}_{2}^{ \pm}+2}}\right]$, The PDF of $\gamma_{2}$ is given by
$
f\left(\gamma_{2}\right)=\frac{1}{\bar{\gamma}_{2}} \exp \left(-\frac{\gamma_{2}}{\bar{\gamma}_{2}}\right) 
$
and
$\bar{\gamma}_{2}=\mathbb{E}\left(\gamma_{2}\right)$

\begin{align*}
S & =(-1)^{\left\lfloor\frac{l v^{v_{2}+k-1}}{\sqrt{M_{2}}}\right\rfloor+\lambda_{k, i, 2}}, \quad c_{k}=2-\delta_{k, 1} \\
D_{2}(k, i) & =2^{k-1}-\left\lfloor\frac{i 2^{k-1}}{\sqrt{M_{2}}}+\frac{1}{2}\right\rfloor \\
D_{3}(k, l) & =2^{v_{1}}-\left\lfloor\frac{l}{2^{1-(k-1) \log _{2}\left(\sqrt{M_{1}}-1\right)}}+\frac{1}{2}\right\rfloor 
\end{align*}
Thus, the average BER of user 2 is given by
\begin{equation}
\bar{P}_{U_{2}}=\frac{2}{v_{2}} \sum_{k=1}^{v_{2}} \bar{P}_{b_{2 k}}. 
\end{equation}

\section{Numerical Simulation and Discussion}

\label{simulationResults}
In this section, the proposed H-NOMA method is compared with the conventional T-NOMA scheme. We consider a path loss exponent of $\lambda=4$ between the BS and user $U_{k}$, corresponding to the Nakagami fading environment. Additionally, $N_{0}$ represents the noise power, with a bandwidth of B=1 MHz and 
\begin{equation}
N_{0} = -174 + 10 \log {10} \text{(B) dBm}
\end{equation}
Furthermore, for  $N = 2$, the distances are set as $d_1 = 1000$ m and $d_2 = 400$ m. For $N = 4$, the distances are $d_1 = 500$ m, $d_2 = 400$ m, $d_3 = 150$ m and $d_4 = 100$ m, following the normalized distance parameters presented in \cite{lee2022novel}. The Rayleigh fading channel for the users is defined as
\begin{equation}
h_{d k}\!\!= \!\!\sqrt{d_{k}^{\lambda }} \times\mathcal{N}\!\!\left(\!\!0,\!\! \frac{\sigma_{d k}^{2}}{2}\!\!\right)\!\!+\!j \mathcal{N}\!\!\left(0, \frac{\sigma_{d k}^{2}}{2}\right)\!/ \!\sqrt{2}, k\!\!=\!\!1,2,3,4
\end{equation}
Since this paper focuses on the analysis of BER performance, the power control allocation mechanism is not considered. We consider the scenario outlined in the table below. The study highlights that the energy efficiency and spectral packing capability of QAM make it a widely adopted modulation scheme in the literature \cite{yahya2023error}. The power allocation is selected based on the methodologies presented in \cite{baig2019closed, aldababsa2020bit}.

The researchers conduct a thorough examination of the BER to validate the error performance of the NOMA scheme \cite{yahya2023error}. It is important to note that the scale of the figure is logarithmic.

\subsubsection{Two users scenario}
The BER of a downlink H-NOMA system is analytically derived and validated through Monte Carlo simulations using equations [(14), (16)]. The results are obtained under the assumption of perfect channel estimation and are based on QPSK modulation. If $q_k$ represents the distance between BS and the  $k^{th}$ user, then the channel coefficient is given by
\begin{equation}
g_k = \sqrt{q_k^{-\zeta}} \hbar_k,
\end{equation}

$\hbar_k \sim \mathcal{C N}(0,1)$ denotes the small-scale fading parameter. The noise at each user, $n_k$, follows a complex normal distribution, $n_k \sim \mathcal{C N}(0, N_0)$. User 1 experiences the lowest channel gain ($g_1 < g_2$), while user 2 has the highest channel gain. The channel envelope, $|g_k|$, follows a Rayleigh distribution. 
To mitigate inter-user interference (IUI) and ensure reliable user separation, SIC is employed. This leads to a power allocation strategy that follows the inverse order of channel gains, i.e., $\alpha_1 > \alpha_2$.
The Monte Carlo simulation is conducted using the following parameters: $q_1 = 6.015$, $q_2 = 1$, $\zeta = 2$, $\alpha_1 = 0.7$, $E_b/N_0 = 1/\sigma_{\mathfrak{n}1}^2 = 1/\sigma_{\mathfrak{n}2}^2$, and $M_1 = M_2 = 4$. These settings ensure a fair comparison and accurate evaluation of the system's performance.
Fig. \ref{fig:BER2users} illustrates the simulated and analytical BER results for users 1 and 2, demonstrating strong agreement across all examined signal-to-noise ratio (SNR) levels. This confirms the accuracy of the analytical predictions, as they align precisely with the simulation results. Notably, user 1 requires an SNR at least 14 dB lower than user 2 to achieve a BER of $10^{-3}$. Additionally, Fig. \ref{fig:IMPERFECTCSI1} compares the analytical BER performance of T-NOMA and H-NOMA. The results clearly indicate that the H-NOMA system significantly outperforms the T-NOMA system in mitigating errors, leading to a substantial improvement in BER performance.

\begin{figure}
    \centering
    \includegraphics[width=9cm]
    {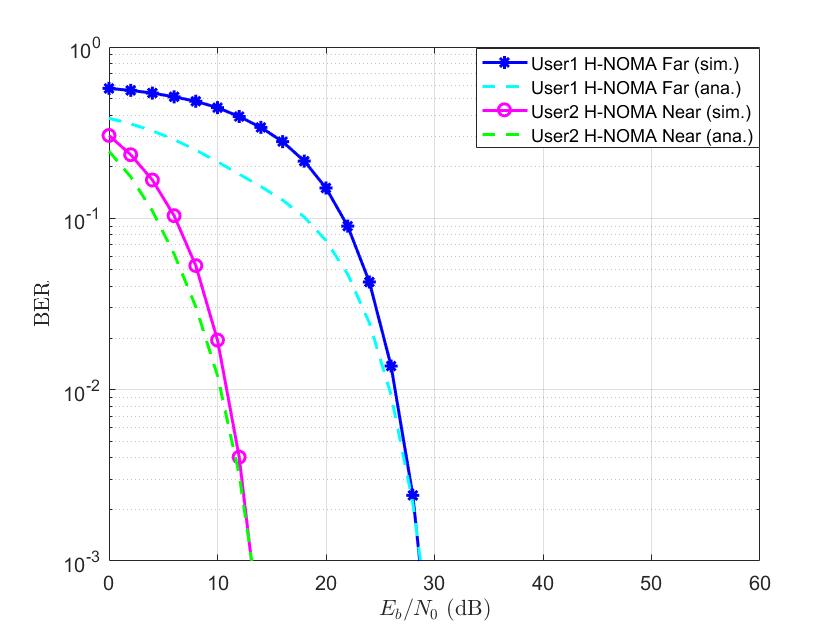}
    \caption{BER analysis for two users H-NOMA.}
\label{fig:BERH-NOMA.}
\end{figure}

\begin{figure}
    \centering
    \includegraphics[width=9cm]
    {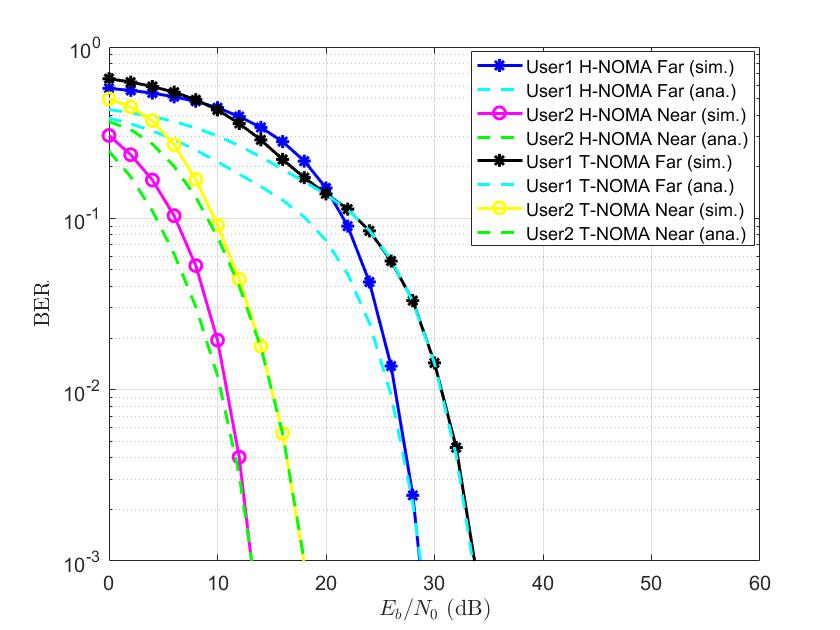}
    \caption{BER analysis and simulation for two users }
\label{fig:BER2users}
\end{figure}
\subsubsection{Four users scenario}

\begin{figure}
    \centering
    \subfloat[]{
        \includegraphics[width=9cm]{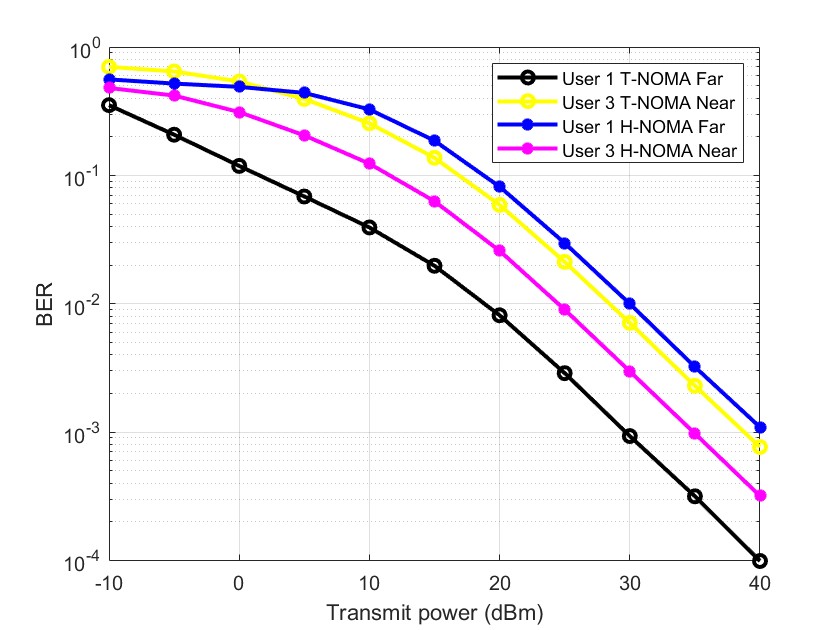}
        \label{fig:Four_users_BER_13}
    }
    \hfill
    \subfloat[]{
        \includegraphics[width=9cm]{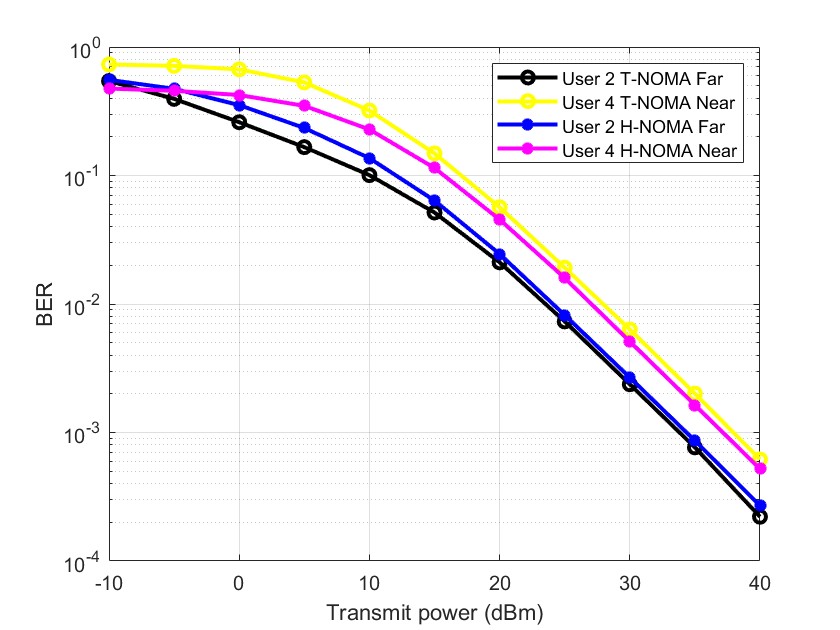}
        \label{fig:Four_users_BER_24}
    }
    \caption{BER comparison for four users.}
    \label{fig:Four_users_BER}
\end{figure}

Examine the scenario depicted in Fig. \ref{fig:Four_users_BER}, where four NOMA users are utilizing 4-QAM, to understand this procedure. The T-NOMA symbols of user 1, who is assigned a greater power coefficient, are represented by the bleu while, the T-NOMA symbols of user 4, who is assigned a lower power coefficient, are represented by the red. we have used the parameters written in table above. The H-NOMA symbols of two user are also depected in the same figure.

At fig \ref{fig:Four_users_BER_13}, it's clear, for the user3 (near), we obtain 5 dB or more performance improvement at a bit-error-rate (BER) of $10^{-3}$ compared to the state-of-the-art existing T-NOMA.
for the user1 (the farest), we obtain a degradation of 7 dB or more at a bit-error-rate (BER) of $10^{-3}$ compared to the state-of-the-art existing T-NOMA.

At fig \ref{fig:Four_users_BER_24}, it's clear, for the user2 (far), we obtain 1 dB or more performance improvement at a bit-error-rate (BER) of $10^{-3}$ compared to the state-of-the-art existing T-NOMA.
for the user4 (the Nearest), we obtain an improvement of 3 dB or more at a bit-error-rate (BER) of $10^{-3}$ compared to the state-of-the-art existing T-NOMA.

So, for user 2,3 and 4, that H-NOMA improves the BER of T-NOMA system for all users of near area and one user of the far area. But, in the case of user 1, T-NOMA shows the best BER performances than H-NOMA. 

The interpretation is that on the contrary of T-NOMA system, H-NOMA system is more resilient to
multipath fading and channel impairments by using Hadamard Transform.

\subsubsection{Imperfect SIC for Two Users Scenario}
T-NOMA provides high spectrum efficiency, but its effectiveness heavily relies on perfect CSI and a fault-free system. CSI errors or system malfunctions can significantly degrade performance, especially for users situated farther from the BS. This degradation is further exacerbated by poor communication channel conditions. Additionally, the number of users that can be served in each SIC cluster is limited because a decoding error in one user will propagate, causing errors in decoding for all subsequent users. Therefore, enhancing the resilience of T-NOMA systems is crucial.

\begin{figure}
    \centering
    \includegraphics[width=9cm]{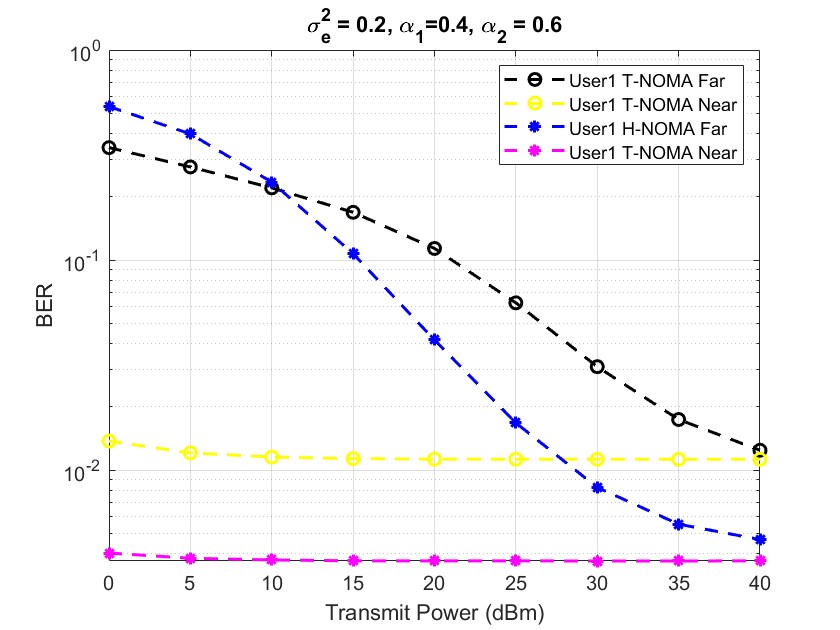}
     \caption{Imperfect Channel Coefficient comparison for two users between T-NOMA and H-NOMA.}
\label{fig:IMPERFECTCSI1}
\end{figure}

\begin{figure} [h]
    \centering
    \includegraphics[width=9cm]{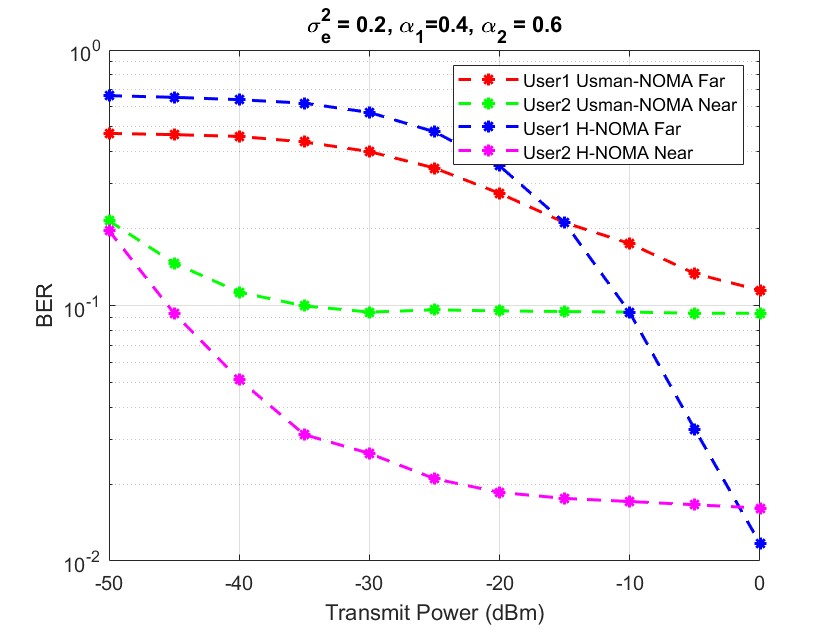}
     \caption{Imperfect Channel Coefficient comparison for two users between Usman-NOMA and H-NOMA.}
\label{fig:IMPERFECTCSI2}
\end{figure}

In Fig. \ref{fig:IMPERFECTCSI1}, it is evident that for the Near user, a performance improvement of 15 dB or more is achieved at a BER of $10^{-2}$ compared to the existing state-of-the-art T-NOMA. For the Far user, a performance improvement of 10 dB or more is observed at a BER of $10^{-1}$ when compared to the existing state-of-the-art T-NOMA. 
In Fig. \ref{fig:IMPERFECTCSI2}, for the Far user, a performance improvement of 15 dB or more is achieved at a BER of $10^{-1}$ compared to the existing Usman-NOMA, where HT is applied after the modulation stage. For the Near user, a clear performance improvement is noted at a base station of 0 dB compared to the existing Usman-NOMA. This suggests that applying HT at the source stage makes the scheme less susceptible to the effects of imperfect CSI errors.

 \subsubsection{Real Time Application}
\label{realTimeApplication}
 
To assess the robustness of the HT-based NOMA scheme (H-NOMA) and compare its performance with the conventional T-NOMA, we conducted simulations using six $512 \times 512$ grayscale test images and varying spectral characteristics, as shown in Fig. 12. For the Monte Carlo simulation and analysis, we used the following parameters: $q_1 = 4$, $q_2 = 3$, $\zeta = 1$ (Rayleigh fading), power allocation factors $\alpha_1 = 0.6$, $E_b/N_0 = 1/\sigma_{\mathfrak{n}1}^2 = 1/\sigma{\mathfrak{n}_2}^2$, $SNR = 25 \text{ dB}$, and $4$-QAM modulation.
To quantify the distortion in the reconstructed data for each user, we employed the peak-to-peak signal-to-noise ratio (PSNR), defined as
\begin{equation}
 PSNR=10 \log_{10}\left ( \frac{255^2}{MSE} \right ),
\end{equation}
where the mean square error (MSE) is given by
\begin{equation}
 MSE=\frac{1}{N\times M}\sum_{i=1}^{N-1}\sum_{j=1}^{M-1}\left ( K\left ( i,j \right )-D \left ( i,j \right ) \right )^2,
\end{equation}
where $N$ and $M$ represent the height and width of the image, respectively. Here, $D$ denotes the estimated image, and $K$ is the original image. We utilized both the standard T-NOMA and the proposed H-NOMA to measure the PSNR of the reconstructed images. The results are presented in Table 3. The table demonstrates that the H-NOMA scheme yields reconstructed images of superior quality compared to the T-NOMA method. Notable differences in the peak-to-peak signal-to-noise ratios (PSNRs) between users 1 and 2 are observed when comparing the two approaches. Specifically, H-NOMA improves the image quality for user 1 (the far user) by over 6 dB, while for user 2 (the near user), H-NOMA achieves an even more significant improvement, increasing the quality by over 17 dB. 
\hspace{-5mm}
\begin{table}[]
\caption{PSNR (dB) values for the images transmitted by T-NOMA and H-NOMA.}
\label{demo-table}
\resizebox{8cm}{!}{
\begin{tabular}{lllll}
\!\!\!\! Test Images &  \!\!\!\! T-NOMA && \!\!\!\!  H-NOMA  \\
\!\!\!\! & User 1 \!\!\!\! & User 2 \!\!\!\!& User 1\!\!\!\! & User 2 \\
 Zelda/Baboon&15.36  &17.47  & 21.84 &35.37  \\
 Goldhill/Barbara& 15.38 & 17.61 &21.97  & 35.43
 \\
 Pepper/Girlface& 15.41 & 17.57 &20.84  & 34.09
\end{tabular}}
\end{table}

\section{Conclusion}
\label {conclusion}

The proposed HT-based NOMA system incorporates HT at the source level, prior to modulation, ensuring compatibility with multiple description coding (MDC) theory. This application of HT disperses the signal, promoting balanced communication between near and far users, which in turn enhances signal quality, accelerates transmission, and reduces the need for retransmissions. By improving the signal-to-noise ratio (SNR) and lowering the bit error rate (BER), HT boosts the system's resilience against hardware impairments, channel state information (CSI) errors, and fading channels.

HT’s signal spreading also improves quality of service (QoS), lowers outage probability, and increases data rates by leveraging frequency diversity to mitigate noise and fading. These results highlight the potential of HT-based NOMA for future communication systems, such as 6G, where reliability and high data rates are paramount. Our findings demonstrate that H-NOMA outperforms T-NOMA, especially in terms of delivering higher-quality image transmissions, as evidenced by superior peak-to-peak signal-to-noise tatio (PSNR) values.


\ifCLASSOPTIONcaptionsoff
  \newpage
\fi

	\bibliographystyle{IEEEtran}
	\bibliography{IEEEabrv,mybibliography}

\begin{thebibliography}{10}
\providecommand{\url}[1]{#1}
\csname url@samestyle\endcsname
\providecommand{\newblock}{\relax}
\providecommand{\bibinfo}[2]{#2}
\providecommand{\BIBentrySTDinterwordspacing}{\spaceskip=0pt\relax}
\providecommand{\BIBentryALTinterwordstretchfactor}{4}
\providecommand{\BIBentryALTinterwordspacing}{\spaceskip=\fontdimen2\font plus
\BIBentryALTinterwordstretchfactor\fontdimen3\font minus \fontdimen4\font\relax}
\providecommand{\BIBforeignlanguage}[2]{{%
\expandafter\ifx\csname l@#1\endcsname\relax
\typeout{** WARNING: IEEEtran.bst: No hyphenation pattern has been}%
\typeout{** loaded for the language `#1'. Using the pattern for}%
\typeout{** the default language instead.}%
\else
\language=\csname l@#1\endcsname
\fi
#2}}
\providecommand{\BIBdecl}{\relax}
\BIBdecl

\bibitem{dai2015non}
L.~Dai, B.~Wang, Y.~Yuan, S.~Han, I.~Chih-Lin, and Z.~Wang, ``Non-orthogonal multiple access for 5g: solutions, challenges, opportunities, and future research trends,'' \emph{IEEE Communications Magazine}, vol.~53, no.~9, pp. 74--81, 2015.

\bibitem{liu2017nonorthogonal}
Y.~Liu, Z.~Qin, M.~Elkashlan, Z.~Ding, A.~Nallanathan, and L.~Hanzo, ``Nonorthogonal multiple access for 5g and beyond,'' \emph{Proceedings of the IEEE}, vol. 105, no.~12, pp. 2347--2381, 2017.

\bibitem{10615971}
M.~Kulhandjian, H.~Kulhandjian, G.~K. Kurt, and H.~Yanikomeroglu, ``Delay-doppler domain pulse design for otfs-noma,'' in \emph{2024 IEEE International Conference on Communications Workshops (ICC Workshops)}, 2024, pp. 63--68.

\bibitem{beddiaf2022unified}
S.~Beddiaf, A.~Khelil, F.~Khennoufa, F.~Kara, H.~Kaya, X.~Li, K.~Rabie, and H.~Yanikomeroglu, ``A unified performance analysis of cooperative noma with practical constraints: Hardware impairment, imperfect sic and csi,'' \emph{IEEE Access}, vol.~10, pp. 132\,931--132\,948, 2022.

\bibitem{xie2022robust}
H.~Xie and Y.~Xu, ``Robust resource allocation for noma-assisted heterogeneous networks,'' \emph{Digital Communications and Networks}, vol.~8, no.~2, pp. 208--214, 2022.

\bibitem{wu2019video}
J.~Wu, B.~Tan, J.~Wu, and M.~Wang, ``Video multicast: Integrating scalability of soft video delivery systems into noma,'' \emph{IEEE Wireless Communications Letters}, vol.~8, no.~6, pp. 1722--1726, 2019.

\bibitem{dani2022resource}
M.~N. Dani, D.~K. So, J.~Tang, and Z.~Ding, ``Resource allocation for layered multicast video streaming in noma systems,'' \emph{IEEE Transactions on Vehicular Technology}, vol.~71, no.~11, pp. 11\,379--11\,394, 2022.

\bibitem{tang2020mdc}
T.~Tang, A.~Wang, S.~Muhaidat, S.~Li, M.~Li, and J.~Liang, ``Mdc-noma: Multiple description coding-based nonorthogonal multiple access for image transmission,'' \emph{IEEE Systems Journal}, vol.~15, no.~3, pp. 3632--3641, 2020.

\bibitem{usman2018joint}
M.~R. Usman, A.~Khan, M.~A. Usman, and S.~Y. Shin, ``Joint non-orthogonal multiple access (noma) \& walsh-hadamard transform: Enhancing the receiver performance,'' \emph{China Communications}, vol.~15, no.~9, pp. 160--177, 2018.

\bibitem{8441999}
I.~Baig, U.~Farooq, N.~Ul~Hasan, M.~Zghaibeh, U.~M. Rana, M.~Imran, and M.~Ayaz, ``On the papr reduction: A novel filtering based hadamard transform precoded uplink mc-noma scheme for 5g cellular networks,'' in \emph{2018 1st International Conference on Computer Applications \& Information Security (ICCAIS)}, 2018, pp. 1--4.

\bibitem{bouslam2024noma}
K.~A. Bouslam, J.~Amadid, F.-Z. Bennioui, R.~Iqdour, and A.~Zeroual, ``Noma based walsh hadamard transform and user pairing strategy,'' in \emph{2024 International Conference on Global Aeronautical Engineering and Satellite Technology (GAST)}.\hskip 1em plus 0.5em minus 0.4em\relax IEEE, 2024, pp. 1--6.

\bibitem{berrouche2014improved}
Y.~Berrouche and R.~E. Bekka, ``Improved multiple description wavelet based image coding using hadamard transform,'' \emph{AEU-International Journal of Electronics and Communications}, vol.~68, no.~10, pp. 976--982, 2014.

\bibitem{1091168}
J.~Pearl, H.~Andrews, and W.~Pratt, ``Performance measures for transform data coding,'' \emph{IEEE Transactions on Communications}, vol.~20, no.~3, pp. 411--415, 1972.

\bibitem{borran2009design}
M.~J. Borran, A.~Sabharwal, and B.~Aazhang, ``Design criterion and construction methods for partially coherent multiple antenna constellations,'' \emph{IEEE transactions on Wireless Communications}, vol.~8, no.~8, pp. 4122--4133, 2009.

\bibitem{assaf2020exact}
T.~Assaf, A.~J. Al-Dweik, M.~S. El~Moursi, H.~Zeineldin, and M.~Al-Jarrah, ``Exact bit error-rate analysis of two-user noma using qam with arbitrary modulation orders,'' \emph{IEEE Communications Letters}, vol.~24, no.~12, pp. 2705--2709, 2020.

\bibitem{lee2022novel}
H.~Y. Lee and S.~Y. Shin, ``A novel user grouping in phase rotation based downlink noma,'' \emph{IEEE Access}, vol.~10, pp. 27\,211--27\,222, 2022.

\bibitem{yahya2023error}
H.~Yahya, A.~Ahmed, E.~Alsusa, A.~Al-Dweik, and Z.~Ding, ``Error rate analysis of noma: Principles, survey and future directions,'' \emph{IEEE Open Journal of the Communications Society}, 2023.

\bibitem{baig2019closed}
S.~Baig, U.~Ali, H.~M. Asif, A.~A. Khan, and S.~Mumtaz, ``Closed-form ber expression for fourier and wavelet transform-based pulse-shaped data in downlink noma,'' \emph{IEEE Communications Letters}, vol.~23, no.~4, pp. 592--595, 2019.

\bibitem{aldababsa2020bit}
M.~Aldababsa, C.~G{\"o}ztepe, G.~K. Kurt, and O.~Kucur, ``Bit error rate for noma network,'' \emph{IEEE Communications Letters}, vol.~24, no.~6, pp. 1188--1191, 2020.

\end{thebibliography}

%

\end{document}